\documentclass[a4paper,12pt,english]{article}

\usepackage{amsfonts,bm,amssymb,euscript,array,babel,cite}

\usepackage{amsmath,amsthm}
\usepackage[dvips]{graphicx}
\usepackage[T1]{fontenc}

\makeatletter

\newcommand*{\dittostraight}{---\textquotedbl---}

\newcommand{\dd}{\mathrm{d}}

\newcommand{\w}{\wedge}
\newcommand{\bbm}{\left(\begin{matrix}}
\newcommand{\ebm}{\end{matrix}\right)}
\newcommand{\beq}{\begin{eqnarray}}
\newcommand{\eeq}{\end{eqnarray}}
\newcommand{\T}{\text{T}}
\newcommand{\M}{\text{M}}
\makeatother

\newcommand{\sfrac}[2]{{\textstyle\frac{#1}{#2}}}

\newcommand{\be}{\begin{equation}}
\newcommand{\ee}{\end{equation}}
\newcommand{\beqa}{\begin{eqnarray}}
\newcommand{\eeqa}{\end{eqnarray}} 

\newcommand{\barr}{\begin{array}}
\newcommand{\earr}{\end{array}}
\numberwithin{equation}{section}
\newcommand{\mf}{\mathfrak}

\def\a{\alpha}  
  
 \def\d{\delta} 

\def\l{\lambda}

\def\R{{\mathbb R}}  
\def\Z{{\mathbb Z}} \def\one{\mbox{1 \kern-.59em {\rm l}}}

\def\bea{\begin{eqnarray}} 
\def\eea{\end{eqnarray}}
\def\nn{\nonumber} 

\def\bit{\begin{itemize}} 
\def\eit{\end{itemize}} 

\def\({\left(} 
\def\){\right)}

\allowdisplaybreaks[3]
\begin{document}

\makeatother


\parindent=0cm
\renewcommand{\title}[1]{\vspace{10mm}\noindent{\Large{\bf
#1}}\vspace{8mm}} \newcommand{\authors}[1]{\noindent{\large
#1}\vspace{5mm}} \newcommand{\address}[1]{{\itshape #1\vspace{2mm}}}


\begin{titlepage}

\begin{flushright}
ITP-UH-20/13
\end{flushright}

\begin{center}

\vskip 3mm

\title{ {\LARGE 
Dirac structures on nilmanifolds \\[6mm] 
and coexistence of fluxes}}

\vskip 3mm

\authors{\large 
Athanasios {Chatzistavrakidis}$^{~\star,}\footnote{thanasis@itp.uni-hannover.de}$, 
Larisa Jonke$^{~\dagger,}\footnote{larisa@irb.hr}$ and 
Olaf Lechtenfeld$^{~\star,}$\footnote{lechtenf@itp.uni-hannover.de}}

\vskip 3mm

\address{  {$^{\star}$\
}Institut f\"ur Theoretische Physik and
Riemann Center for Geometry and Physics,  \\
Leibniz Universit\"at Hannover,  Appelstra{\ss}e 2, 30167 Hannover, Germany 

\

{$^{\dagger}$}\
Theoretical Physics Division, 
Rudjer Bo$\check s$kovi\'c Institute, \\
Bijeni$\check c$ka 54, 10000  Zagreb, Croatia
}

\bigskip 

\end{center}

\vskip 2cm

\textbf{Abstract}

\vskip 3mm


We study some aspects of the generalized geometry of nilmanifolds and examine 
to which extent different types of fluxes can coexist on them. Nilmanifolds constitute 
a class of homogeneous spaces which are interesting in string compactifications 
with fluxes since they carry geometric flux by construction. They are generalized 
Calabi-Yau spaces and therefore simple examples of generalized geometry at work. 
We identify and classify Dirac structures on nilmanifolds, which are maximally isotropic 
subbundles closed under the Courant bracket. In the presence of non-vanishing fluxes, 
these structures are twisted and closed under appropriate extensions of the Courant bracket.
Twisted Dirac structures on a nilmanifold may carry multiple coexistent fluxes of any type. 
We also show how dual Dirac structures combine to Courant algebroids and work out an 
explicit example where all types of generalized fluxes coexist. These results may be useful 
in the context of general flux compactifications in string theory.

%


\end{titlepage}

\tableofcontents

\section{Introduction} 

Fluxes are an important ingredient of all modern approaches to string compactifications \cite{FC}. In particular they serve as a 
powerful tool in the efforts to reveal interesting string vacua and stabilize their moduli. They conventionally appear as 
vacuum expectation values of antisymmetric tensor fields or as non-trivial geometric twists \cite{SS,kstt}. The existence of more general types of fluxes 
was indicated in the study of unconventional string backgrounds \cite{kstt} and studied in \cite{Shelton:2005cf,Dabholkar:2005ve}, but for some time their precise geometric description remained unclear. 
Such fluxes were dubbed ``non-geometric'' but progress in their understanding has yielded this term a misnomer 
and therefore it will not be used in the present paper. 
Indeed, these more general flux types should be understood via a generalized approach to geometric structures.

Generalized Complex Geometry (GCG) is a generalization of standard complex and symplectic geometry \cite{gcg1,gcg2}, which 
has also provided new tools for string compactifications \cite{gcgsugra}. It is in fact the appropriate framework to study the 
generalized fluxes that were mentioned above without introducing mathematically obscure additional 
ingredients. The essence of this setting lies in the organization and extension of the geometric transformations of the fields of 
a theory, i.e. diffeomorphisms and gauge transformations, to O$(\text d,\text d)$ transformations. The latter is identical to 
the T-duality group of string theory compactified on a d-dimensional torus\footnote{A related but independent development, implementing  T-duality at the level of (extended) target space, goes under the name of double field theory, see refs.\cite{dft} for recent reviews.}. 

In this paper we utilize certain tools of GCG to study fluxes on nilmanifolds. The latter are simple GC manifolds, see ref. \cite{sixmanifolds}, which 
carry a natural geometric flux and they have appeared in numerous instances as internal spaces in string compactifications, 
e.g. in refs. \cite{SS,nil}. 
Here we pose the following main question:
\begin{itemize}
 \item How much can we dress a nilmanifold with fluxes?
\end{itemize}
or, equivalently,
\begin{itemize}
 \item Can all types of fluxes coexist on a nilmanifold?
\end{itemize}
Let us already stress that this question does not refer to cases which are T-dual to purely geometric ones in the 
standard sense \cite{Dall'Agata:2007sr}. The question has an affirmative answer in certain such cases but this is not a remarkable result from the viewpoint of new possibilities for string vacua. 
Here we examine such a possibility when a geometric dual is not available.

As a strategic choice we use the notion of Dirac structure in GCG \cite{dirac}. Dirac structures are maximally isotropic subbundles of 
a (twisted) Courant algebroid which are closed under the (twisted) Courant bracket. This makes them attractive because 
the Jacobi identity is always satisfied on them and in a sense one can define a good coordinate system on these 
submanifolds. Concentrating on step-2 nilmanifolds{\footnote{Higher-step nilmanifolds can be studied with the same 
tools. This choice is here just a simplification which can be raised if necessary.}}, we study and classify the Dirac 
structures on them by considering arbitrary deformations introduced by tensor fields of rank $(p,q)$ with $p+q=2$. These 
include 2-forms $B$, not necessarily closed, 2-vectors $\beta$, not necessarily Poisson and mixed (1,1) tensors. The Dirac structures 
which close under the Courant bracket introduce the integrability conditions of closed $B$ and Poisson $\beta$ and they 
come in two general types, spanned by a specific form of generalized vectors. These are fluxless cases corresponding to 
constant background moduli. In particular, in the 3-dimensional example of the Heisenberg nilmanifold the moduli 
space of Dirac structures is 5-dimensional, consisting of one three-parameter family and one two-parameter family.
More possibilities arise when deformed Courant brackets are considered. In particular, introducing an $H$-twisted 
Courant bracket, where $H=\dd B$, allows for twisted Dirac structures which are closed under the new bracket. 
Similarly an $R$-twisted bracket, where $R=\sfrac 12[\beta,\beta]_S$, the latter being the Schouten bracket, lead to 
twisted Dirac structures under it. These cases correspond to turning on $H$ and $R$ flux respectively, on top of the 
already present geometric $f$ flux. The general setting of these results was essentially described before, 
e.g. in refs. \cite{Halmagyi:2008dr,Blumenhagen:2012pc,Szabo}, albeit without reference to Dirac structures. Twisted 
Poisson structures in string theory were also discussed in ref. \cite{Lust:2012fp}.

The most interesting aspect comes about when both $H$ and $R$ fluxes are considered simultaneously. This naturally leads to 
a bracket with both $H$ and $R$ twists, a particular version of the Roytenberg bracket\cite{Roy}. We examine the conditions under which 
Dirac structures are obtained, thus showing that they may carry multiple fluxes of any type. In particular, it turns 
out that apart from the geometric flux $f$, present by construction, $H$, $R$ and also $Q$ fluxes appear. A 
detailed example based on the Heisenberg nilmanifold is presented where  we identify all the non-vanishing components of the 
$H,R,f$ and $Q$ fluxes as well as their origin.

Moreover we discuss how two Dirac structures are combined to form a (twisted) Lie bialgebroid. 
The latter provides a Courant algebroid when the two Dirac structures are orthogonal, or dual, to each other 
\cite{wein,severa}. In the 
fluxless case we show that it is possible to combine the two families of Dirac structures on a step-2 nilmanifold 
such that a Courant algebroid with all moduli fields present is obtained. In particular, for the 3-dimensional 
Heisenberg nilmanifold all five parameters are non-vanishing. This is achieved by rotating appropriately the bases 
of the two structures such that they become orthogonal. 
The rotation is performed with an element of type (1,1) which is a contraction  $B\cdot \beta$. 
This method can be also extended to the cases involving multiple fluxes. 
In particular, we show  that 
a twisted Courant algebroid with all flux types in coexistence can be constructed on the Heisenberg nilmanifold. 
The associated bracket is again the one with $H$ and $R$ twist 
and the two dual almost Dirac structures are closed under this bracket.  

These results  provide an affirmative answer to the main question that we posed 
above, showing that flux backgrounds without geometric duals are in principle possible. Although we do not deal here 
with the question of whether this mechanism  actually leads to true string backgrounds, we suggest our findings as an 
indication that a lot more possibilities exist for flux backgrounds  than the ones that have been studied before. 
In particular we would expect that such constructions can provide in the long run an explanation for the origin 
of all 4-dimensional gauged supergravities.

\section{Generalized geometry of step-2 nilmanifolds} 

\subsection{Brief review of generalized geometry}

In this brief section we collect some definitions and results of generalized geometry. In particular, we are going to 
present only the material which is necessary for the comprehension of the rest of the paper. A more detailed presentation
may be found in ref. \cite{gcg2}.

Generalized geometry \cite{gcg1,gcg2} extends the standard tangent bundle over a manifold M to the sum of its tangent and cotangent bundles, $\T\M\oplus \T^\ast \M$, at least locally. Sections of this vector bundle are generalized vectors, $\mf X=X+\eta,\;X\in \T\M,\;\eta\in \T^\ast \M$. 
One can define  the Courant bracket on the sections of
$\T\M\oplus \T^{\ast}\M$ by the formula
\beq\label{cour}[X+\eta,Y+\xi]_C=[X,Y]_{L}+{\cal L}_{X}\xi-{\cal L}_{Y}\eta-\frac{1}{2}\dd (\iota_X\xi-\iota_Y\eta),\eeq
where we used the standard Lie bracket for vectors, $[\;,\;]_{L}$, the Lie derivative of 1-forms along vectors, ${\cal L}_X\eta=\iota_X\dd \eta+\dd(\iota_X\eta)$, and the contraction of vectors and forms, $\iota_X\eta$. 
Adding a smooth map  $\rho:\T\M\oplus \T^{\ast}\M\to \T\M$,
called anchor, plus a bilinear form
\beq\label{bilinear}\langle X+\eta,Y+\xi\rangle=\frac{1}{2}(\iota_X\xi+\iota_Y\eta),\eeq
one obtains (with some additional compatibility conditions between the aforementioned structures) a Courant algebroid.
The bracket (\ref{cour}) is skew-symmetric, but does not satisfy the Jacobi identity\footnote{
One could instead define the Dorfman bracket $$ [X+\eta,Y+\xi]_D=[X,Y]_{L}+{\cal L}_{X}\xi-\iota_Y\dd\eta,$$ which does satisfy the Jacobi identity, but it fails to be  skew-symmetric.}.
In particular, the Jacobiator of three
generalized vectors $\mathfrak X, \mathfrak Y, \mathfrak Z$ under the Courant bracket is given by
\beq
{\rm Jac}(\mf X,\mf Y,\mf Z)=[[\mf X,\mf Y]_C,\mf Z]_C+{\rm cycl.}=\dd N(\mf X,\mf Y,\mf Z),
\eeq
where $N$ is the Nijenhuis operator, defined as
\beq
3 N(\mf X,\mf Y,\mf Z)=\langle[\mf X,\mf Y]_C,\mf Z\rangle+\langle[\mf Y,\mf Z]_C,\mf X\rangle+ \langle[\mf Z,\mf X]_C,\mf Y\rangle.
\eeq

A Courant algebroid can alternatively be constructed out of Lie bialgebroids via a
 generalization of the Drinfeld double construction \cite{wein}. One starts from a Lie algebroid,  a vector bundle $L$ over a manifold M equipped with a Lie bracket $\{\;,\;\}$ and an anchor map $\rho:L\rightarrow \T\M$, which on the space $\Gamma(L)$ of sections of $L$ satisfies
\beq
&&\rho(\{X,Y\})=\{\rho(X),\rho(Y)\},\;X,Y\in\Gamma(L)\label{an},\\[6pt]
&&\{X,fY\}=f\{X,Y\}+({\cal L}_{\rho(X)}f)Y, \;f\in C^{\infty}(\textrm{M}). 
\eeq
 Then one  defines a Lie bialgebroid
as a pair of Lie algebroid structures on a vector bundle $L$ and its dual, $L^\ast$,  having a  unique  extension  to a Courant algebroid structure on $L\oplus L^\ast$, with the symmetric
form of the type (\ref{bilinear}). For the original Courant bracket, $L = \T\M$ with the usual bracket of vector fields and the anchor being the identity, and
$L^\ast=\T^\ast\M$ with the zero bracket on 1-forms and a zero anchor. Note that when the manifold M is reduced to a point,  one recovers Lie (bi)algebras. 

In this paper we focus on a specific type of Lie algebroids, the Dirac structures. We have seen that the Jacobiator of the Courant bracket is proportional to an exact term of a bilinear expression. 
We can then define an almost Dirac structure as a maximally isotropic subbundle $L$ of a Courant algebroid satisfying $\langle X+\eta,Y+\xi\rangle=0,\forall ~ X+\eta, Y+\xi\in L$, on which the Jacobiator vanishes. If this subbundle is involutive, i.e. closed under the Courant bracket restricted to $L$, we obtain an integrable Dirac structure (obviously, a Lie algebroid). As an example of a Dirac structure one can take the tangent bundle, $\T\M\subset \T\M\oplus \T^\ast\M$. This subbundle is isotropic, i.e., $\langle X,Y\rangle=0,\forall ~ X, Y\in \T\M$. Moreover, the restriction of the Courant bracket to $\T\M$ is the standard Lie bracket of vector fields and thus involutive. 
The  construction of explicit Dirac structures as well as of some additional structures within the framework of generalized geometry will be discussed in the following sections for M being a nilmanifold.

\subsection{Nilmanifolds}

Nilmanifolds are homogeneous manifolds which incorporate geometric fluxes and  are constructed as orbits of a lattice in a nilpotent Lie group. The nilpotence degree of the underlying Lie group is transfered to the nilpotence index or step of the associated nilmanifold. This index can be read off from the  relations between the structure constants of the  associated Lie algebra.
The invariant 1-forms which span the cotangent bundle of an arbitrary step-2 nilmanifold in d dimensions
can be written as
\be\label{forms1}
e^i=\d^i_c\dd x^c-\sfrac 12 f^i_{bc}x^b\dd x^c=e^i_c\dd x^c,
\ee
where $x^a$ are Cartesian coordinates and $f_{ab}^i$ are antisymmetric 
in their lower indices. These are the structure constants of a step-2 nilpotent Lie algebra and they satisfy the relation 
\be \label{int1}f_{ai}^{c}f_{bc}^{j}=0,\ee 
with no summation over repeated indices. 
Indices $a,b,c,\dots$ are used for the coordinate basis, spanned by the $\dd x^a$, while 
indices $i,j,k,\dots$ are reserved for the so-called Mal'cev basis, spanned by $e^i$.
These 1-forms satisfy the Maurer-Cartan equations
\be \label{mc}
\dd e^i+\sfrac 12 f^{i}_{jk}e^j\w e^k=0
\ee
and, unlike the $\dd x^a$, they are globally well defined.
The invariant 1-vectors which span the tangent bundle of a nilmanifold are given as
\be \label{vectors1}
\theta_i=\d_i^c\partial_c-\sfrac 12 f_{ib}^{c}x^b\partial_c=\tilde e_i^c\partial_c,
\ee
and they are dual to the 1-forms $e^i$. Indeed, eq. (\ref{bilinear}) implies that 
\be
\langle\theta_i,e^j\rangle=\sfrac 12 \d^j_i.
\ee
In the above expressions we introduced the twist matrix $e^i_c$, which is defined as
\be 
e^i_c=\d^i_c-\sfrac 12f^i_{bc}x^b,
\ee
as well as its inverse 
\be
\tilde e_i^c=\d_i^c+\sfrac 12f_{ai}^{c}x^a.
\ee
These two quantities satisfy the obvious relations
\bea
e^i_c\tilde e^c_j&=&\d^i_j, \\
\tilde e^c_ie^i_d&=&\d^c_d.
\eea
Moreover, an obvious relation between the structure constants and the twist matrix is
\be
f^i_{ab}=2\partial_{[b}e^i_{a]}. 
\ee
The above 1-forms
and 1-vectors can be initially defined on the covering space of the
nilmanifold, which is a group manifold, and then they are projected to the homogeneous space under the action of a discrete subgroup. 
This generalizes the corresponding construction of a torus as a quotient of $\R^d$ by $\Z^d$. The identifications that have 
to be made take the following general form,
\be 
(x^a,\partial_b)\sim (x^a+c^a+\sfrac 12 f^a_{jb}c^jx^b,\partial_b+\sfrac 12 f^a_{bk}c^k\partial_a),
\ee
where $c^i\in 2\pi R^i \Z$ and $R^i$ the radii of the corresponding cycles.

For a step-2 nilmanifold, the Maurer-Cartan equations take also the alternative form
\be 
\dd e^i=-\sfrac 12f^i_{bc}\dd x^b\wedge \dd x^c.
\ee
Comparing with (\ref{mc}) we read off the relation $f^i_{jk}=f^i_{bc}\tilde e^b_j\tilde e^c_k$, which on step-2 manifold becomes $f^i_{jk}=f^i_{bc}\delta^b_j\delta^c_k$.
The simplicity of the above geometric quantities allows for a simple set of closed formulae for the interior product 
of vectors and forms as well as for the corresponding Lie derivatives. Indeed this set of equations for the interior 
product is
\be
\begin{aligned}
\iota_{\partial_a}\dd x^b&=\delta_a^b,\\
\iota_{\theta_i}\dd x^b&=\d_i^b-\sfrac 12 f_{ic}^bx^c=\tilde e_i^b,\\
\iota_{\partial_a}e^j&=\d_a^j+\sfrac 12f^j_{ac}x^c=e_a^j,\\
\iota_{\theta_i}e^j&=\d_i^j.
\end{aligned}
\ee
The corresponding Lie derivatives may be easily calculated, and they give the following expressions,
\be 
\begin{aligned} 
{\cal L}_{\partial_a}\dd x^b&=0, \\
{\cal L}_{\theta_i}\dd x^b&=-\sfrac 12 f_{ij}^be^j,\\
{\cal L}_{\partial_a}e^j&=-\sfrac 12 f_{ai}^je^i,\\
{\cal L}_{\theta_i}e^j&=-f_{ik}^je^k.
\end{aligned}
\ee
The above expressions are useful in explicit computations.

The sections of the direct sum of the tangent and the cotangent bundle,
$\T\M\oplus \T^{\ast}\M$, are arbitrary 
generalized vectors
\be 
\mf X=u_ie^i+v^i\theta_i,
\ee
$u_i, v^i$ being constant coefficients. This is essentially an expansion over the basis 
of the extended bundle. In particular, $\T\M$ is d-dimensional with basis $\{\theta_i\}$, $\T^{\ast}\M$ is also d-dimensional
with basis $\{e^i\}$ and $\T\M\oplus\T^{\ast}\M$ is 2d-dimensional. The natural basis of the latter is given by the 
trivial extension of the bases of its constituents, namely by the 2d generalized vectors $\{\theta^i+0,0+e^i\}$. Note that 
we used the same symbol to denote the zero 1-vector and the zero 1-form but this should cause no confusion to the careful reader.

The Lie bracket of two 1-vectors $\theta_i$ in the Mal'cev basis is 
\be 
[\theta_i,\theta_j]_L=f_{ij}^{k}\theta_k.
\ee 
Using the properties of the Lie bracket and the Lie derivative, the Courant bracket is computed as
\be 
[\mf X,\mf Y]_C=v^i\tilde v^j f_{ij}^k\theta_k + (v^i\tilde u_j-\tilde v^i u_j)f_{ki}^je^k \equiv \l^k\theta_k+\mu_ke^k :=\mf Z,
\ee
where $\mf X=u_ie^i+v^i\theta_i$ and $\mf Y=\tilde u_ie^i+\tilde v^i\theta_i$. This makes evident the fact
that the Courant bracket is a closed 
operation.

\section{Dirac structures on nilmanifolds}

\subsection{Deformations and Dirac structures}

Although the Courant bracket is a closed operation for any maximal set of generalized vectors, this ceases to be 
generally true for a non-maximal subset of generalized vectors on the Courant algebroid. 
However, one can find rank d subbundles where the bracket does close. These are exactly the Dirac structures on the manifold 
\cite{dirac}.
For a nilmanifold one sees immediately that both $\T\M$ and $\T^\ast \M$ are  Dirac structures, since both the 
tangent and the cotangent bundles are isotropic, i.e.\ the bilinear form (\ref{bilinear}) vanishes for 1-vectors and 1-forms separately; on both subspaces the Courant bracket is closed, with the only non-vanishing one being $[\theta_i,\theta_j]_C=[\theta_i,\theta_j]_L=f_{ij}^{k}\theta_k,\;\theta_i\in \T\M$ and the anchor maps being identity and zero for $\T\M$ and $\T^\ast \M$, respectively. Let us denote the subbundles with basis elements $\theta_i$ and $e^i$ as $L_0$ and $L_0^{\ast}$ respectively, namely
\be 
(L_0)_i:=\theta_i, \quad (L_0^{\ast})^i:=e^i.
\ee
Then obviously the bialgebroid $L_0\oplus L_0^{\ast}$ is a Courant algebroid, since according to eq. (\ref{bilinear}) 
$\langle \theta_i,e^j\rangle=\sfrac 12 \d^j_i$.

It is important to realize that the above subbundles are not the only ones which constitute Dirac structures. One can work towards 
a classification of Dirac structures on a nilmanifold by considering arbitrary deformations. The possible deformations are generated by tensors of rank $(p,q), p+q=2$, with $(2,0)$ tensors being  2-vectors, not necessarily Poisson, $(0,2)$ tensors being a 2-forms, not necessarily closed, and  $(1,1)$ tensors being mixed.
This can be seen as follows. The general element of the Lie algebra $\mathfrak{so}(L\oplus L^\ast)\simeq \mathfrak{so}(\text{d},\text{d})$
acting on $L\oplus L^\ast$ can be decomposed as \bea \begin{pmatrix}F & \beta   \\    B & -F^\ast \end{pmatrix},\;F\in\textrm{End} (L),\; B:L\rightarrow L^\ast, \;\beta: L^\ast\rightarrow L. \nn\eea
Starting from $L=\T\M$ we see that $B$ is 2-form, $\beta$ is 2-vector and the endomorphism $F:L\rightarrow L$ and its dual $F^\ast:L^\ast\rightarrow L^\ast$ can be represented by a $(1,1)$ mixed tensor.
Exponentiation of the general algebra element produces an orthogonal transformation which leaves the bilinear form (\ref{bilinear}) invariant\cite{gcg2}. For example, 
\bea\label{expbeta}
e^\beta:=\begin{pmatrix}1 & \beta   \\    0 & 1 \end{pmatrix},\;\;
e^\beta(X+\eta)=X+\iota_X\beta+\eta.\nn\eea
In the following we will use the symmetry transformations generated by $\exp(B)$, $\exp(\beta)$, and $\exp(F)$ to deform the Dirac structures\footnote{These transformations were called twist transformation in \cite{DA09}.}. 

Note that any d-dimensional nilmanifold could be thought of as the plain torus $T^{\text{d}}$ with ``geometric flux'' turned on.
Indeed, let us start from a plain torus, which is also a step-1 nilmanifold. 
The corresponding generalized vectors are
$\mf X=v^i\partial_i+ u_i\dd x^i$.   
Acting on the generalized vectors with the mixed tensor\footnote{The wedge product for the mixed tensors is defined as $\dd x^i\wedge \partial_j=\dd x^i\otimes\partial_j - \partial_j\otimes\dd x^i.$ 
It does not antisymmetrize the upper and lower indices.} $F=\frac{1}{2}f_{ab}^{c}x^a\dd x^b\wedge\partial_c$ we obtain:
\beq\label{geomf}
e^F \mf X=v^i(\d_i^c\partial_c+\frac{1}{2}f_{ai}^{c}x^a\partial_c)+ u_i(\d^i_c\dd x^c-\frac{1}{2}f_{ab}^{i}x^a\dd x^b)
=v^i\theta_i+ u_ie^i,\eeq
where $e^i$ and $\theta_i$ are given by the expressions (\ref{forms1}) and  (\ref{vectors1}), respectively.
Higher-order terms vanish in this case because of the condition (\ref{int1}),
which holds identically for any step-2 nilmanifold. This is no longer true for higher-step nilmanifolds.
The expression (\ref{int1}) can be recast in the form of a bracket as 
\be \label{int2}[F,F]_S=0,\ee
where the generalized Schouten bracket $[\cdot,\cdot]_S$ is the extension of the Courant bracket for higher-rank generalized vectors and defined as
\bea\label{fcond}
&&[\mf X_1\wedge\cdots\wedge \mf X_p,\mf Y_1\wedge\cdots\wedge\mf Y_q]_S= \\[4pt]
&&~~~~~~~~=\sum_{ij}(-1)^{i+j}
[\mf X_i,\mf Y_j]_C\wedge \mf X_1\wedge\cdots\wedge\hat {\mf X}_i\wedge\cdots\wedge \mf X_p\wedge\mf Y_1
\wedge\cdots\wedge\hat{\mf Y}_j\wedge\cdots\wedge\mf Y_q,\nn\eea
where the hat over a generalized vector denotes exclusion.
The Dirac structures of the case under study remain isotropic and integrable exactly due to the condition (\ref{int1}), or 
equivalently (\ref{int2}). In that sense, these expressions should be thought of as integrability conditions for almost 
Dirac structures.

Let us reconsider a step-2 nilmanifold as a starting point with the standard basis of 1-vectors $\theta_i$ and 
globally well-defined 1-forms $e^i$. We would now like to deform the corresponding Dirac structures
using an arbitrary 2-form $B$ or 2-vector $\beta$, 
which can be $x$-dependent quantities and are written as
\bea 
&&B=\sfrac 12 \tilde B_{ab}(x)\dd x^a\wedge \dd x^b=\sfrac 12  B_{ij}(x)e^i\wedge e^j,\\[6pt]
&&\beta=\sfrac 12 \tilde\beta^{ab}(x)\partial_a\wedge \partial_b=\sfrac 12 \beta^{ij}(x)\theta_i\wedge \theta_j,
\eea
where we wrote two expressions, one for each basis. The relation among the 
two sets is given by the equations
\bea
&&\tilde B_{cd}= B_{ij}e^i_ce^j_d,\\[6pt]
&&\tilde \beta^{cd}= \beta^{ij}\tilde e_i^c\tilde e_j^d.
\eea

There are two cases to be examined, in particular
\bea \label{singledef1}
&&(L_B)_i:=e^{-B}(L_0)_i=\tilde e_i^c(\partial_c-\tilde B_{cl}\dd x^l),\\[6pt]
&&(L^{\ast}_{\beta})^i:=e^{\beta}(L_0^{\ast})^i=e^i_c(\dd x^c+\tilde\beta^{cl}\partial_l),\label{singledef4}
\eea
in obvious notation, where the subscript 
denotes the type of the deformation. The other two possibilities are trivial, i.e. 
$(L_{\beta})_i:=e^{\beta}\theta_i=\theta_i\equiv (L_0)_i$ and 
$(L^{\ast}_{B})^i:=e^{-B}e^i=e^i\equiv (L^{\ast}_0)^i.$ In other words, the $\theta_i$ subbundle is stable under 2-vector deformations, while the 
$e^i$ one is stable under 2-form deformations. The expressions (\ref{singledef1}) and (\ref{singledef4}) were 
written in terms of the coordinate basis, but it is more useful to express them in the Mal'cev basis. 
Indeed, in this basis they simply become
\bea \label{singledef1b}
&&(L_B)_i=\theta_i- B_{ij}e^j,\\[6pt]
&&(L^{\ast}_{\beta})^i=e^i+\beta^{ij}\theta_j.\label{singledef4b}
\eea
In the following we will use the latter expressions, i.e. we will stick to the globally well-defined elements of the 
Mal'cev basis.

Next, we would like to know which of the above possible deformations are still integrable, i.e. under which ones the bracket remains closed so that they still constitute Dirac structures. From the physical point of view, we would like to know which is the 
maximal possible set of fluxes (or background moduli in the constant case)
that  could be turned on over the nilmanifold as a compactification manifold in a string-theory framework. 
In order to examine this question, let us exploit the expressions (\ref{singledef1b})-(\ref{singledef4b}).

\paragraph{Isotropy.} 
The first rather trivial step is to examine the isotropy of the structures under the bilinear form.
This is straightforward. It holds that
\bea 
\langle L_B,L_B\rangle=\langle L^{\ast}_{\beta},L^{\ast}_{\beta}\rangle=0,
\eea
where the indices where suppressed. This is true for any deformation $B$ and $\beta$ respectively, and therefore no 
additional requirements are introduced. This renders $L_{B}$ and $L^{\ast}_{\beta}$ almost Dirac structures.

\paragraph{Closure.}
The second criterion concerns the closure of the Courant bracket on the subbundle, which would upgrade each almost 
Dirac structure to a Dirac structure indeed. 
The Courant bracket of two $L_B$ elements is found to be
\be \label{vbc}
[(L_B)_i,(L_B)_j]_C=f_{ij}^{k}(L_B)_k- 3(\theta_{[k}B_{ij]}+f^{m}_{[ki}B_{j]m})e^k,
\ee
where we repeatedly used the integrability condition (\ref{int1}). In order for the Courant bracket 
to close, a sufficient condition is that the second term vanishes. This is true as long as the 2-form $B$ has vanishing exterior derivative, namely
\be 
\dd B=0 \qquad \Rightarrow \qquad \mbox{closure of $L_B$ Courant bracket} .
\ee
Indeed, recall that the components $B_{ij}$ correspond to the Mal'cev basis and therefore 
\be
\begin{aligned} 
\dd B&=\sfrac 12 \dd B_{ij}\w e^i\w e^j+\sfrac 12 B_{ij}\dd (e^i\w e^j)\\
&=\sfrac 12 \partial_{a}B_{ij}\dd x^a\w e^i\w e^j-\sfrac 14B_{ij}(f^i_{kl}e^k\w e^l\w e^j-f^{j}_{kl}e^k\w e^l\w e^i)\\
&=\sfrac 16(3\partial_a B_{ij}\tilde e^a_ke^i\w e^j\w e^k-3B_{ij}f^i_{kl}e^j\w e^k\w e^l)\\
&=\sfrac 16 3(\theta_kB_{ij}+f^m_{ki}B_{jm})e^i\w e^j\w e^k,
\end{aligned}
\ee
which proves the above assertion.
This does not mean that the 2-form has to be constant. Indeed, it is enough to have, for example,
\be 
 B_{ab}=c^{[a}x^{b]}+\text{constant}.
\ee
Then this proves that 
\be 
\dd  B=0 \qquad \Rightarrow \qquad \mbox{$L_B$ is a Dirac structure} .
\ee

On the other hand, for the second case we compute
\be\label{fbetac}
[(L^{\ast}_{\beta})^i,(L^{\ast}_{\beta})^j]_C=(\theta_k\beta^{ij}+2f^{[j}_{\ kl}\beta^{i]l})(L^{\ast}_{\beta})^k+
3(\beta^{l[k}(\theta_l\beta^{ji]})+\beta^{l[i}\beta^{\underline{m}j}f^{k]}_{ml})\theta_k.
\ee
In order for this bracket to close we find 
\be 
[\beta,\beta]_S=0 \qquad \Rightarrow \qquad \mbox{closure of $L^{\ast}_{\beta}$ Courant bracket} .
\ee
In other words, $ \beta$ better be a Poisson 2-vector respecting the geometric twist. Note once more that we work in 
the Mal'cev basis, where
\be 
[\beta,\beta]_S=\sfrac 16 3(\beta^{lk}\theta_l\beta^{ji}+\beta^{li}\beta^{{m}j}f^{k}_{ml})\theta_i\w \theta_j\w\theta_k.
\ee
The above show that 
\be 
[\beta,\beta]_S=0 \qquad \Rightarrow \qquad \mbox{$L^{\ast}_{\beta}$ is a Dirac structure} .
\ee

\paragraph{Anchor maps.}
The definition of a Lie algebroid includes an anchor map compatible with the bracket on the algebroid as in eq. (\ref{an}).
For the Dirac structures that we discussed above the anchor maps are given as
\bea 
\rho&:& L_B\rightarrow e^{-B}\T\M,\;\quad\rho(\theta_i-B_{ij}e^j)=\theta_i-B_{ij}e^j,\label{an1}\\[6pt]
\rho^\ast&:& L^\ast_\beta\rightarrow e^{-B}\T\M,\;\quad\rho^\ast(e^i+\beta^{ij}\theta_j)=\beta^{ij}(\theta_j-B_{jk}e^k). \label{an2}
\eeq

Summarizing, on an arbitrary step-2 nilmanifold there are two families of 
Dirac structures, and the results appear in the following table.
\begin{center}\boxed{
\begin{tabular}{cccc}
 \underline{Dirac structure} & \underline{Basis} &\underline{Bracket} & \underline{Condition} 
 \\[4pt]
 $L_B$& $\Theta_i:=\theta_i- B_{ij}e^j$ & $[~,~]_C$ & $\dd  B=0$
 \\[4pt]
 $L^{\ast}_{\beta}$ & $E^i:=e^i+\beta^{ij}\theta_j$& $[~,~]_C$ & $[\beta,\beta]_S=0$
\end{tabular}}
\end{center}

\subsection{Twisted Dirac structures and fluxes}

Although in the beginning of this section we considered general 2-form and 2-vector deformations,
the closure of the Courant bracket was very 
restrictive. It led to the conditions of $B$ being closed and $\beta$ being Poisson. In physical terms, with the definitions 
\be
H=\dd B \quad \mbox{and}\quad R=\sfrac 12[\beta,\beta]_S,
\ee 
these conditions mean that there is no $H$ or $R$ flux respectively.

The inclusion of fluxes in the present framework is rather straightforward. 
For example, for the NS-NS flux $H$, the question now becomes whether an integrable deformation for $H=\dd B\neq 0$ can be introduced.
In this case  one  defines the $H$-twisted Courant bracket:
\beq\label{hbracketa}
[X+\eta,Y+\xi]_H=[X+\eta,Y+\xi]_C+\tau_H,
\eeq
where
\be
\tau_H=\iota_Y\iota_X H.
\ee
We can now define a $H$-twisted Dirac structure which is closed under the $H$-twisted Courant bracket for $\dd H=0$ \cite{gcg2}. 
Indeed, in the $L_{B}$ case we directly compute
\be
{(\tau_H)}_{ij}=3(\theta_{[k}B_{ij]}+f^{m}_{[ki}B_{j]m})e^k.
\ee
Then it directly follows from eq. (\ref{vbc}) that 
\be
[(L_B)_i,(L_B)_j]_H=f_{ij}^{k}(L_B)_k,
\ee
which shows that the $H$-twisted bracket automatically closes without any further conditions. Here we use the identity as the anchor map compatible with $H$-twisted bracket.

A similar strategy is followed for the case of $R$ flux, which corresponds to the case of the 2-vector $\beta$ 
not being Poisson. The important role here is played by the Schouten bracket for $\beta$. As we mentioned before, when $\beta$ is Poisson it satisfies
\be
[\beta,\beta]_S=0.
\ee
However, in general the Schouten bracket gives a 3-vector 
$R=\sfrac 16R^{ijk}\theta_i\w\theta_j\w\theta_k=\sfrac 12 [\beta,\beta]_S$.
It is then natural to use an alternative bracket, the Roytenberg or simply $R$-bracket.
This is defined as 
\be 
[X+\eta,Y+\xi]_R=[X+\eta,Y+\xi]_C-\tau_R,
\ee
where now
\be
\tau_R=R(\eta,\xi,\cdot).
\ee
In particular, in the $L^{\ast}_{\beta}$ case we directly compute
\be
(\tau_R)^{ij}=3(\beta^{l[k}(\theta_l\beta^{ji]})+\beta^{l[i}\beta^{\underline{m}j}f^{k]}_{ml})\theta_k.
\ee
Then it is evident from eq. (\ref{fbetac}) that
\be
[(L^{\ast}_{\beta})^i,(L^{\ast}_{\beta})^j]_R=Q^{ij}_{k}(L^{\ast}_{\beta})^k,
\ee
where 
\be\label{q}
Q^{ij}_{k}=\theta_k\beta^{ij}+2f^{[j}_{kl}\beta^{i]l}.
\ee
This shows that the $R$-twisted bracket closes automatically without further conditions, and the $L^{\ast}_{\beta}$ is a 
$R$-twisted Dirac structure. The anchor map compatible with the $R$-twisted bracket is the one given in eq. (\ref{an2}).

A less obvious result is obtained by attempting to use the $R$-twisted bracket in the case of $L_B$ and asking 
whether closure of the almost twisted Dirac structure under this bracket is possible. 
One then computes
\be 
(\tau_R)_{ij}=B_{il}B_{jm}(\beta^{n[l}\theta_n\beta^{km]}+\beta^{n[m}\beta^{\underline{p}l}f^{k]}_{pn})\theta_k.
\ee
Then the $R$-bracket takes the form
\be \label{vbr}
\begin{aligned}
{[}(L_B)_i,(L_B)_j]_R &= f_{ij}^{k}(L_B)_k- 3(\theta_{[k}B_{ij]}+f^{m}_{[ki}B_{j]m})e^k\\
&\quad -B_{il}B_{jm}(\beta^{n[l}\theta_n\beta^{km]}+\beta^{n[m}\beta^{\underline{p}l}f^{k]}_{pn})\theta_k.
\end{aligned}
\ee
It is then observed that this bracket closes under the condition
\be\label{conditionVB}
3(\theta_{[r}B_{ij]}+f^{m}_{[ri}B_{j]m})=-B_{il}B_{jm}B_{kr}
(\beta^{n[l}\theta_n\beta^{km]}+\beta^{n[m}\beta^{\underline{p}l}f^{k]}_{pn}).
\ee
Indeed, then the result is 
\be \label{vbr2}
[(L_B)_i,(L_B)_j]_R=f_{ij}'^{k}(L_B)_k,
\ee
with 
\be
f_{ij}'^{k}=f_{ij}^{k}-B_{il}B_{jm}(\beta^{n[l}\theta_n\beta^{km]}+\beta^{n[m}\beta^{\underline{p}l}f^{k]}_{\ pn}).
\ee
Therefore, the $R$-bracket for $L_B$ closes under the condition (\ref{conditionVB}), and then $L_B$ is a 
$R$-twisted Dirac structure. We will discuss the meaning of this statement in a while. The anchor map in this case is also the identity map.

Similarly there is a further interesting possibility obtained by evaluating the $H$-twisted 
Dirac bracket for the almost Dirac structure $L^{\ast}_{\beta}$. First of all, one computes
\be
(\tau_H)^{cd}=\beta^{ci}\beta^{dj}(\theta_{[k}B_{ij]}+f^{m}_{[ki}B_{j]m})e^k.
\ee
Then the $H$-twisted bracket becomes 
\be\label{fbetah}
\begin{aligned}
{[}(L^{\ast}_{\beta})^i,(L^{\ast}_{\beta})^j]_H&=Q^{ij}_{k}(L^{\ast}_{\beta})^k+
3(\beta^{l[k}(\theta_l\beta^{ji]})+\beta^{l[i}\beta^{\underline{m}j}f^{k]}_{ml})\theta_k\\
&\quad +\beta^{ip}\beta^{jq}(\theta_{[k}B_{pq]}+f^{m}_{[kp}B_{q]m})e^k,
\end{aligned}
\ee
which closes only under the  condition
\be\label{conditionFb}
3(\beta^{l[r}(\theta_l\beta^{ji]})+\beta^{l[i}\beta^{\underline{m}j}f^{r]}_{ml})=
\beta^{ip}\beta^{jq}\beta^{kr}(\theta_{[k}B_{pq]}+f^{m}_{[kp}B_{q]m}).
\ee
If satisfied, then we obtain
\be\label{fbetah2}
[(L^\ast_{\beta})^i,(L^\ast_{\beta})^j]_H=Q'^{ij}_{k}(L^\ast_{\beta})^k
\ee
with
\be \label{qprime}
Q'^{ij}_{k}=Q^{ij}_{k}+\beta^{ip}\beta^{jq}(\theta_{[k}B_{pq]}+f^{m}_{[kp}B_{q]m}).
\ee
Therefore, the $H$-twisted bracket closes for $L^{\ast}_{\beta}$ under the condition (\ref{conditionFb}). This result was previously obtained in an analysis of twisted Poisson structures performed in~\cite{ks,sw}. In this case the anchor map compatible with the $H$-twisted bracket is given as in (\ref{an2}).

Finally, one could ask  what happens with the almost Dirac structures $L_B$ and $L^\ast_\beta$ when we deform  the bracket both with $H$ and $R$ twists. 
We evaluate explicitly   the following extended bracket 
\be\label{hrbra}
[\mathfrak X,\mathfrak Y]_{H R}=[\mathfrak X,\mathfrak Y]_{C}+\tau_H-\tau_R.
\ee
One can easily show that 
\bea \label{vbrh}
[(L_B)_i,(L_B)_j]_{HR}=f_{ij}^{k}(L_B)_k-B_{il}B_{jm}(\beta^{n[l}\theta_n\beta^{km]}+\beta^{n[m}\beta^{\underline{p}l}f^{k]}_{pn})\theta_k,
\eea
so that the integrability condition for the $H$- and $R$-twisted almost Dirac structure $L_B$ is 
\be
(\tau_R)_{ij}\in L_B  \qquad\Leftrightarrow\qquad 
B_{il}B_{jm}B_{kn}(\beta^{p[l}\theta_p\beta^{nm]}+\beta^{q[m}\beta^{\underline{p}l}f^{n]}_{pq})=0\ .
\ee
For the dual structure we obtain
\bea\label{fbetahr}
[(L^{\ast}_{\beta})^i,(L^{\ast}_{\beta})^j]_{HR}&=&Q^{ij}_{k}(L^{\ast}_{\beta})^k+\beta^{ip}\beta^{jq}(\theta_{[k}B_{pq]}+f^{m}_{[kp}B_{q]m})e^k,
\eea
so the integrability condition in this case is 
\be
(\tau_H)^{ij}\in L^\ast_\beta \qquad\Leftrightarrow\qquad 
\beta^{ip}\beta^{jq}\beta^{kr}(\theta_{[r}B_{pq]}+f^{m}_{[rp}B_{q]m})=0\ .
\ee
The anchor maps for the last two cases will be analyzed in section 4.

Summarizing, the resulting possibilities for twisted Dirac structures on step-2 nilmanifolds are given in the following table:
\begin{center}\boxed{
\begin{tabular}{clc}
 \underline{Twisted Dirac structure} & \underline{Bracket} & \underline{Condition} 
 \\[4pt]
 $L_B$ & $[~,~]_H$ & -
 \\[4pt]
 $L^{\ast}_{\beta}$ & $[~,~]_R$ & - 
 \\[4pt]
 $L_B$ & $[~,~]_R$ & $H_{ijk}=-\sfrac 13 B_{il} B_{jm} B_{kn}R^{lmn}$
 \\[4pt]
 $L^{\ast}_{\beta}$ & $[~,~]_H$ &  $R^{ijk}=\sfrac 13\beta^{il}\beta^{jm}\beta^{kn}H_{lmn}$
\\[4pt]
$L_B$ & $[~,~]_{HR}$ & $ B_{il}B_{jm}B_{kn}R^{lmn}=0$
\\[4pt]
$L^{\ast}_{\beta}$ & $[~,~]_{HR}$ & $\beta^{il}\beta^{jm}\beta^{kn}H_{lmn}=0$
\end{tabular}}
\end{center}

The following remarks are in order. First of all, it is easy to see that in the middle two 
cases the $H$ and $R$ fluxes are interrelated. This means that in case the one on the 
right hand side of any equation vanishes, so does the other. 
Such cases essentially boil down either to the previous ones of untwisted Dirac structures or to Dirac 
structures with different fluxes along different cycles of the manifold (when the manifold is of 
sufficiently high number of dimensions, e.g. six). This is interesting but not remarkable because 
one can always find a geometric dual of these set-ups.
However, the last two lines are much less restrictive and much more interesting.
Indeed, in these cases  both types of $H$ and $R$ flux can be simultaneously and independently present
in the same Dirac structure under a well-defined twist of the 
bracket. 
Therefore, we encounter cases where a geometric dual is not available. 
The doubly twisted bracket plays an instrumental role in this construction. This will become clear in the 
example that follows in Subsection 3.3.3.
Finally, let us note that all these results refer only to individual Dirac structures. Whether these can be consistently 
 combined to form Lie bialgebroids and therefore Courant algebroids is a different issue, which we address 
 in Section 4.

\subsection{Example: Dirac structures on the Heisenberg nilmanifold}

Let us now apply the above results to the simplest possible 
case of the 3-dimensional Heisenberg nilmanifold. This will make more clear 
the amount of parameters in the two families of Dirac structures, as well as 
the fact that our statements about the cases where fluxes are present 
are meaningful and not empty.

 For the case at hand, let us choose the unique non-vanishing structure constant to be 
 $f_{12}^{3}=-f_{21}^{3}=1$, and pick the polarization where the 
globally well-defined 1-forms are
 \be \label{hef}
 e^1=\dd x^1, \quad e^2=\dd x^2, \quad e^3=\dd x^3-\sfrac 12 x^1\dd x^2+\sfrac 12
 x^2\dd x^1\ .
 \ee
 Hence, the dual 1-vectors are
 \be \label{hev}
 \theta_1= \partial_1-\sfrac 12 x^2\partial_3, \quad \theta_2=\partial_2+\sfrac
 12 x^1\partial_3, \quad \theta_3=\partial_3.
 \ee
 Then the Courant bracket of two arbitrary generalized vectors reads explicitly as
 \be
 [\mf X,\mf Y]_C=(v^1\tilde v^2-v^2\tilde v^1)\theta_3+(v^2\tilde u_3-\tilde v^2 u_3)e^1-(v^1\tilde u_3-\tilde v^1 u_3)e^2.
 \ee
 The elements $e^1$, $e^2$ and $\theta_3$ are all central. This implies that the Courant bracket always produces a central element.

\subsubsection{Constant moduli - fluxless case}

Let us now examine the classification of Dirac structures. As we discussed 
before, there are two such families accompanied by two conditions. 
The families have the general form 
\bea  
\Theta_i&=&\theta_i- B_{ij}e^j,\\[6pt]
E^i&=&e^i+\beta^{ij}\theta_j,
\eea  
with $i,j=1,2,3$. Therefore, before imposing the integrability 
conditions there are six parameters in total, three for each family. 
Turning to the conditions, the first one is that the 2-form $B=\sfrac 12 B_{ij}e^i\wedge e^j$ is closed.
For constant moduli $B_{ij}$ this is satisfied automatically due to the fact that 
\be 
\dd (e^i\wedge e^j)=0.
\ee 
Therefore no reduction of the parameters occurs for the first family, 
which is a genuine three-parameter one. 
The second condition is that $\beta=\sfrac 12 \beta^{ij}\theta_i\wedge 
\theta_j$ is a Poisson 2-vector. For constant moduli $\beta^{ij}$ we 
compute
\be
\begin{aligned} 
{[}\beta,\beta]_S &=\beta^{ij}\beta^{kl}f_{ik}^{m}\theta_m\wedge\theta_j\wedge\theta_l\\[4pt]
&=2(\beta^{12})^2\theta_1\wedge \theta_2\wedge\theta_3.\label{Rheisenberg}
\end{aligned}
\ee
This vanishes only for $\beta^{12}=0$. Therefore we conclude that 
a reduction of the number of 
parameters by one occurs for the second family of Dirac structures, thus 
leaving only a two-parameter family. These two families, with a total of 
five parameters, exhaust the Dirac structures on the Heisenberg nilmanifold.

\subsubsection{Single twist deformations}

The above result changes for twisted Dirac structures, i.e.
 in the presence of fluxes. In that case all six parameters of the two families 
can be retained. Let us work out in some detail the corresponding twisted 
cases. 

Consider the 2-form
\be
B=\sfrac N3x^1e^2\wedge e^3+\sfrac N3x^2e^3\wedge e^1+\sfrac N3x^3e^1\wedge e^2,
\ee
whose exterior derivative is
\be 
H=\dd B=Ne^1\wedge e^2\wedge e^3.
\ee
This obviously satisfies $\dd H=0$, as it should. Then the subbundle $L_B$, spanned by $\Theta_i$, is a twisted Dirac structure 
under the $H$-twisted bracket without any further restrictions.

In the previous section we claimed that $L_B$ can be an $R$-twisted Dirac structure as well, under a 
condition that relates $H$ and $R$. Examining this condition in the present case,
it boils down to $N=0$ for any $R$. This means that 
the above 2-form $B$ is forced to vanish. More generally, $B$ just needs to be closed. 
In other words, $L_B$ is also an $R$-twisted Dirac structure as long as $\dd B=0$, i.e.~the $H$ flux vanishes. 
This is the $H{=}0$ limit of the doubly-twisted Dirac structure appearing in the fifth line of the previous table.

Similar results hold for the second type of Dirac structure. Indeed, one can invoke (\ref{Rheisenberg}) to introduce 
a non-Poisson 2-vector. It is enough to consider the previous one, i.e. 
\be 
\beta = \sfrac12 \beta^{ij}\theta_i\wedge\theta_j
\ee
with constant parameters, albeit 
without imposing the restriction of vanishing $\beta^{12}$.
Then 
\be
R=(\beta^{12})^2\theta_1\wedge \theta_2\wedge\theta_3,
\ee
and $L^{\ast}_{\beta}$, spanned by $E^i$, is a twisted Dirac structure under the $R$-twisted bracket without any 
further restrictions.

As before, we should check whether an $H$-twisted Dirac structure is obtained as well. This is indeed the case, however 
the restriction $\beta^{12}=0$ enters again. This means that $L^{\ast}_\beta$ is also an $H$-twisted Dirac structure if $[\beta,\beta]_S=0$, i.e.~in the absence of $R$ flux. 
This is the $R{=}0$ limit of the doubly-twisted Dirac structure appearing in the sixth line of the previous table.

\subsubsection{Multiple twist deformations}

In order to analyze the case with both twists, we consider the following set-up of 2-form and 2-vector,
\bea 
B&=&Nx^1e^2\w e^3,\label{e1b}\\[4pt]
\beta&=&\sqrt{c}\theta_1\w\theta_2,\label{e1beta}
\eea
where $\sqrt{c}=\beta^{12}$ is some constant. The corresponding fluxes are given as 
\bea 
H&=&Ne^1\w e^2\w e^3,\\[4pt]
R&=&c\theta_1\w\theta_2\w\theta_3.
\eea 
As a consequence,
\bea 
\dd H&=&0, \\[4pt]
{[}\beta,R]_S&=&0,
\eea
which are obvious because there is no 4-form or 4-vector on a 3-dimensional manifold.
The associated Dirac structures are
\bea 
L_B&=&\{\theta_1,\theta_2-Nx^1e^3,\theta_3+Nx^1e^2\}\ =\ \{\Theta_i\},\\[4pt]
L^{\ast}_{\beta}&=&\{e^1+\sqrt{c}\theta_2,e^2-\sqrt{c}\theta_1,e^3\}\ =\ \{E^i\}.
\eea
These Dirac structures are both  $H$- and $R$-twisted, and one can easily check that the conditions from  Subsection 3.2 are fulfilled for  $B$ and $\beta$ given in (\ref{e1b}) and (\ref{e1beta}), respectively. 
We evaluate explicitly   the  extended bracket (\ref{hrbra}), starting from the contributions from the twists,
\bea\label{tw11}
&(\tau_H)_{ij}=N\epsilon_{ijk}e^k,\quad &(\tau_H)^{ij}=cN\epsilon^{ij3}e^3,\\[4pt]
&(\tau_R)^{ij}=c\epsilon^{ijk}\theta_k,\quad &(\tau_R)_{ij}=c(Nx^1)^2\epsilon_{1ij}\theta_1,
\eea
where 
\bea
(\tau_H)^{ij} &=& \beta^{ik}\beta^{jl} (\tau_H)_{kl},\\[4pt]
(\tau_R)_{ij} &=& B_{ik} B_{jl} (\tau_R)^{kl}.
\eea
For $L_B$ with the $HR$-twisted bracket we find
\be\label{ThetaCR1}
\begin{aligned}
{[}\Theta_1,\Theta_2]_{HR}&=\theta_3+Nx^1e^2=\Theta_3,\\
{[}\Theta_2,\Theta_3]_{HR}&=-c(Nx^1)^2\Theta_1,\\
{[}\Theta_1,\Theta_3]_{HR}&=0.
\end{aligned}
\ee
It is observed that the bracket closes, at least with some non-constant coefficients. 
Similarly, the $HR$-twisted bracket for $L^{\ast}_{\beta}$ yields
\be \label{ECR1}
\begin{aligned}
{[}E^1,E^2]_{HR}&=cNE^3,\\
{[}E^2,E^3]_{HR}&=\sqrt{c}E^2,\\
{[}E^1,E^3]_{HR}&=\sqrt{c}E^1.
\end{aligned}
\ee

Even in this simple example of the 3-dimensional nilmanifold, the $HR$-twisted Dirac structures carry a 
plethora of fluxes of all types. We postpone a detailed presentation of the components and the origin of these fluxes 
to Subsection 4.2.2, after we will have discussed the construction of Courant algebroids based on these Dirac structures.

\section{Lie bialgebroids from dual Dirac structures}

\subsection{Construction of bialgebroids}

Let us now examine whether we can combine the  Dirac and twisted Dirac structures that were identified in the previous 
section into Lie bialgebroids defining 
Courant and twisted Courant algebroids. This is done by combining two Dirac structures $L$ and $L^{\ast}$ as 
$L\oplus L^{\ast}$, such that they are orthogonal with respect to the bilinear form (\ref{bilinear}), namely 
\be \label{ortho}
\langle (L)_i,(L^{\ast})^j\rangle=\sfrac 12 \d^j_i.
\ee
This means that these mutually dual Dirac structures must satisfy the same relation as the elementary ones $L_0$ and $L_0^{\ast}$, for 
which $\langle \theta_i,e^j\rangle=\sfrac 12 \d^j_i$ according to the definition (\ref{bilinear}).

We consider the most general possibility. To construct a Lie bialgebroid, we attempt to combine the 
two general classes of Dirac structures on a nilmanifold, i.e.
\be
\mathfrak L=L_B\oplus L^{\ast}_{\beta}\ .
\ee
To establish that $\mathfrak L$ is a Courant or twisted Courant algebroid, 
it remains to check the orthogonality condition (\ref{ortho}).
We directly compute
\be
\sfrac12 G_i^j:=\langle (L_B)_i,(L^{\ast}_{\beta})^j\rangle=\sfrac 12(\d^j_i+ B_{ik}\beta^{kj})=\sfrac 12(\d^j_i+F^j_i),
\ee
where we defined the specific tensor of (1,1) type\footnote{
The important role of such elements 
may be also appreciated by looking at the fluxes appearing in \cite{Aldazabal:2011nj}, 
although those were obtained in a different context.}
\be
F=B\cdot\beta=B_{ik}\beta^{kj}e^i\wedge\theta_j.
\ee
In general, the scalar product matrix $G_i^j$ is 
not equal to $\d^j_i$. There are two ways to proceed. The first one demands that 
$ B_{ik}\beta^{kj}=0$. However, we shall see that this is too restrictive. Here we will 
follow the second, more general and more interesting path.

Recall that any $\text{O}(\text d,\text d)$ transformation $M$ of (1,1) type, acting as $(L_B)_i\to(M\,L_B)_i$ and
$(L^{\ast}_{\beta})^i\to(M^{-t}L^{\ast}_{\beta})^i$, where $-t$ denotes the inverse transpose, does not alter the matrix $G$ of bilinear 
products. However, in combining two Dirac structures, there is the freedom of choosing bases in each one independently of the other one, by defining
\bea
( L_B)'_i&=&C_i^{\ j}(L_B)_j, \\[6pt]
( L_{\beta}^{\ast})'^i&=&D^i_{\ j}(L_{\beta}^{\ast})^j,
\eea
where $C$ and $D$ are arbitrary but unrelated GL(d) matrices. Obviously, these transformations do not change the type of the Dirac structure, since they just rotate the corresponding bases. The matrix of scalar products in the new bases is 
\be
G'^j_i=2\langle (L_B)'_i,(L^{\ast}_{\beta})'^j\rangle=C_i^{\ k}G_k^{\ l}D_l^{\ j}=(CGD^t)_i^j\overset{!}= \d^j_i,
\ee
where in the last step  we demand it to acquire the desired form. Therefore, we are looking for matrices $C$ and $D$ such that 
the following matrix equation is satisfied,
\be \label{CGD}
CGD^t=\one_{\dd}.
\ee
There are many solutions to this equation, as long as $G$ is invertible. For example, one can consider 
\be
\text{(i)}\quad  C=G^{-1}, \quad D=\one_{\dd}
\ee
or 
\be
\text{(ii)}\quad  C=\one_{\dd}, \quad D=G^{-t}.
\ee
A more general solution may be written as 
\be 
C=G^{-\alpha}, \quad D=G^{(\alpha-1)t}.
\ee
We obtain that $\mathfrak L$ is indeed a Lie bialgebroid, and it defines a Courant or twisted Courant algebroid, depending on the bracket and the additional 
conditions. In particular, in the fluxless case, $\mathfrak L$ is a Courant algebroid with the standard, untwisted 
Courant bracket, under the conditions that $\dd  B=0$ and $[\beta,\beta]_S=0$. Note that the present section is  independent of whether $B$ is a closed 2-form and $\beta$ is a Poisson 2-vector. 
This allows us to construct twisted Courant algebroids with several fluxes turned on. 
In the following specific example we will give explicit expressions for $C$ and $D$.

\subsection{Example: Courant algebroids on the Heisenberg nilmanifold}

\subsubsection{Fluxless case}
Let us revisit the simple example of the 3-dimensional Heisenberg nilmanifold from the viewpoint of Lie 
bialgebroids. Since we are in three dimensions we can express the parameters as
\bea 
B_{ik}&=&\epsilon_{ijk}\a^j,\\[6pt]
\beta^{ik}&=&\epsilon^{ijk}\rho_j.
\eea
For the moment we work with constant moduli. 
Thus $B$ is a closed 2-form and $\beta$ a Poisson 2-vector. 
The vanishing of the Schouten bracket demands $\rho_3$ to be zero.

The bases for the two types of Dirac structures then read
\bea 
L_B&:=&\{\theta_i-\epsilon_{ijk}\a^je^k\},\\[6pt]
L^{\ast}_{\beta}&:=&\{e^i+\epsilon^{ijk}\rho_j\theta_k\}.
\eea
It is directly computed that
\be 
G^m_i=2\langle (L_B)_i,(L^{\ast}_{\beta})^m\rangle=\d^m_i(1+\vec \a\cdot\vec\rho)-\a^m\rho_i.
\ee
This matrix can easily be inverted,
\be
(G^{-1})_m^j=\sfrac 1{1+\vec \a\cdot\vec\rho}(\d_m^j+\rho_m\a^j).
\ee
The two extremal solutions to (\ref{CGD}) are thus given by
\bea
\text{(i)}\quad C&=&\frac{1+\vec\rho\otimes\vec\a}{1+\vec\a\cdot\vec\rho}, \quad D=\one_3, \\
\text{(ii)}\quad C&=&\one_3,\quad D=\frac{1+\vec\a\otimes\vec\rho}{1+\vec\a\cdot\vec\rho}
\eea
and lead respectively to
\bea
\text{(i)}\quad \{(L_B)'_i\}&=&
\sfrac 1{1+\vec\a\cdot\vec\rho}\{\theta_i-\epsilon_{ijk}\a^je^k+\rho_i\vec\a\cdot\vec\theta\}, \quad
\{(L^{\ast}_{\beta})^{\prime i}\}=\{e^i+\epsilon^{ijk}\rho_j\theta_k\},\\[6pt]
\text{(ii)}\quad \{(L_B)'_i\}&=&\{\theta_i-\epsilon_{ijk}\a^je^k\},\quad \{(L^{\ast}_{\beta})^{\prime i}\}=
\sfrac 1{1+\vec\a\cdot\vec\rho}\{e^i+\epsilon^{ijk}\rho_j\theta_k+\a^i\vec\rho\cdot\vec e\}.
\eea
Intermediate solutions can also be worked out. In any case, the bialgebroid $L'_B\oplus L'^{\ast}_{\beta}$ is a Courant algebroid with the standard Courant bracket, and all 
the moduli are present; none of them, apart from $\beta^{12}$, has to vanish. The vanishing of $\rho_3=\beta^{12}$ 
guarantees that the Courant bracket in the subbundles $L'_B$ and $L'^{\ast}_{\beta}$ closes and that they remain Dirac structures.

\subsubsection{Multiple coexistent fluxes}

We now turn to the most interesting case when fluxes are 
present in the construction. As we already pointed out, the discussion in Subsection 4.1 is rather general and does not depend on the fluxes. We illustrate the situation in an example.

We start from the setting of Subsection 3.3.3, with 
\be 
B=Nx^1e^2\wedge e^3,\quad\beta=\sqrt{c}\theta_1\wedge\theta_2
\ee 
and $H$- and $R$-twisted  Dirac structures 
\bea 
L_B&=&\{\theta_1,\theta_2-Nx^1e^3,\theta_3+Nx^1e^2\}=\{\Theta_i\}, \label{Thetai}\\[6pt]
L^{\ast}_{\beta}&=&\{e^1+\sqrt{c}\theta_2,e^2-\sqrt{c}\theta_1,e^3\}=\{E_i\}. \label{Ei}
\eea
We have shown that these structures are involutive under the extended bracket (\ref{hrbra}). 
Let us now consider the sum of the two twisted Dirac structures, $\mathfrak L=L_B\oplus L^{\ast}_{\beta}$. 
We only need to find the proper basis rotations $C$ and $D$ to produce a twisted Courant algebroid.
The only non-vanishing combination of $B\cdot\beta$ is $B_{32}\beta^{21}=\sqrt{c}Nx^1\ne 0$. In other words, the matrix $G$ is 
\be 
G=\begin{pmatrix}
1 & 0 & 0 \\
0 & 1 & 0 \\
\sqrt{c}Nx^1 & 0 & 1
\end{pmatrix}.
\ee 
Inverting it is trivial,
\be 
G^{-1}=\begin{pmatrix}
1 & 0 & 0 \\
0 & 1 & 0 \\
-\sqrt{c}Nx^1 & 0 & 1
\end{pmatrix}.
\ee 
We choose here the solution $C=G^{-1}$ and $D=\one_3$. This yields the Dirac 
structures in the rotated bases,
\bea 
L_B&=&\{\Theta'_i\}=\{\theta_1,\theta_2-Nx^1e^3,\theta_3-\sqrt{c}Nx^1\theta_1+Nx^1e^2\},\\[6pt]
L^{\ast}_{\beta}&=&\{E'^i\}=\{e^1+\sqrt{c}\theta_2,e^2-\sqrt{c}\theta_1,e^3\},
\eea
which differs from (\ref{Thetai}) and (\ref{Ei}) only by a shift of $\Theta_3$,
\be 
\Theta'_3 = \Theta_3 -\sqrt{c}Nx^1\Theta_1\ .
\ee
The isotropy of each structure is evidently preserved. 
Finally, let us take a look at the extended brackets in the new basis. For the elements of $L_B$ we have:
\be\label{ThetaCR1b}
\begin{aligned}
{}[\Theta_1,\Theta_2]_{HR}&=\Theta'_3+\sqrt{c}Nx^1\Theta_1,\\
{[}\Theta_2,\Theta'_3]_{HR}&=\sqrt{c}Nx^1\Theta'_3,\\
{[}\Theta_1,\Theta'_3]_{HR}&=-\sqrt{c}N\Theta_1
\end{aligned}
\ee
while the commutation relations for $L^{\ast}_{\beta}$ remain the same as in (\ref{ECR1}).
This proves the assertion that $L_{B}\oplus L^{\ast}_{\beta}$, with the 
twisted Dirac structures in the above bases, is an ${HR}$-twisted Courant algebroid. In this construction we anchor both subbundles into $L_B=e^{-B}\T\M$:
\bea
\rho&:&L_B\rightarrow L_B,\quad \rho(\Theta_i)=\Theta_i,\\[6pt]
\rho^\ast&:& L^\ast_\beta\rightarrow L_B,\quad\rho^\ast(E^i)=\beta^{ij}\Theta_j.
\eea
The anchor $\rho$ is compatible with the bracket on $L_B$, i.e. $\rho([\Theta_i,\Theta_j]_{HR})=[\rho(\Theta_i), \rho(\Theta_j)]_{HR}$, but for $L^\ast_\beta$ we find that
\be
\rho^\ast([E^i, E^j]_{HR})= [\rho^\ast(E^i),\rho^\ast(E^j)]_{HR}+\rho(R(E^i, E^j,\cdot)).
\ee
These anchor maps turn $(L_B,L^\ast_\beta)$ into a quasi-Lie bialgebroid which gives rise to a Courant algebroid structure on $L_B\oplus L^\ast_\beta$ \cite{Roy}.

The importance of this result lies in the fact that all different types of fluxes coexist in the above construction. Before listing them, let us 
compute the mixed ${HR}$-commutators for completeness. These are
\bea 
{[}\Theta_1,E^1]_{HR}&=&\sqrt{c}\Theta'_3-\sqrt{c}Nx^1E^2+\sqrt{c}NE^3,\nn\\
{[}\Theta_1,E^2]_{HR}&=&0,\nn\\
{[}\Theta_1,E^3]_{HR}&=&-E^2-\sqrt{c}\Theta_1,\nn\\
{[}\Theta_2,E^1]_{HR}&=&\sqrt{c}Nx^1E^1,\nn\\
{[}\Theta_2,E^2]_{HR}&=&\sqrt{c}\Theta'_3, \\
{[}\Theta_2,E^3]_{HR}&=&E^1-\sqrt{c}\Theta_2-\sqrt{c}Nx^1E^3,\nn\\
{[}\Theta'_3,E^1]_{HR}&=&-\sqrt{c}NE^1+cN\Theta_2,\nn\\
{[}\Theta'_3,E^2]_{HR}&=&-cN\Theta_1,\nn\\
{[}\Theta'_3,E^3]_{HR}&=&\sqrt{c}Nx^1E^2.\nn
\eea
We group the fluxes that have appeared. 
By construction there are the metric flux $f^3_{12}=1$, which 
defines the nilmanifold, the $H=\dd B$ flux $H_{123}=N$ and the $R=\sfrac 12[\beta,\beta]_S$ flux 
$R^{123}=c$, appearing below.
\begin{center}\boxed{
\begin{tabular}{ccc}
 \underline{Value} & \underline{Flux} & \underline{Origin}
 \\[4pt]
1 & $f^3_{12}$ & Metric\\[4pt]
$N$ & $H_{123}$ & $H=\dd B$ \\[4pt]
$c$ & $R^{123}$ & $R=\sfrac 12[\beta,\beta]_S$
\end{tabular}}
\end{center}
From the sets of commutation relations we see additional 
effective fluxes with the general index structure $F^k_{ij}$ and 
$Q_{i}^{jk}$. These fluxes are not fundamental since they originate
from non-vanishing combinations of the metric, $B$ and $\beta$. 
They are listed in the following table together with their origin.
\begin{center}\boxed{
\begin{tabular}{ccc}
 \underline{Value} & \underline{Fluxes} & \underline{Origin}
 \\[4pt]
$cN$ & $\tilde Q_{3}^{12}$ & $\tilde Q=Q'-Q=\sfrac 12\beta^{il}\beta^{jm}H_{ijn}\theta_l\w\theta_m\w e^n$ \\[4pt]
$\sqrt{c}$ & $ Q_{2}^{23}, Q_1^{13}$ & $ Q=\beta^{il}f_{lm}^{\ \ n}\theta_n\w\theta_i\w e^m$\\[4pt]
$\sqrt{c}N$ & $F^1_{31}$ & $F=\partial_m(B_{nl}\beta^{li})\theta_i\w e^m\w e^n$\\[4pt]
$\sqrt{c}Nx^1$ & $\tilde F^1_{12}$ & $\tilde F=\sfrac 12\beta^{il}B_{lj}f^{j}_{\ mn}\theta_i\w e^m\w e^n$\\[4pt]
$\sqrt{c}Nx^1$ & $\check F^3_{23}$ & $\check F=B_{il}\beta^{lj}f_{jm}^{\ \ n}e^i\w e^m\w\theta_n$
\end{tabular}}
\end{center}
For the first two lines, we refer to the expressions (\ref{q}) and (\ref{qprime}) and we do not exhibit their vanishing terms in 
the example at hand. Quantities with the index structure $F^i_{ij}$ and $Q_i^{ij}$ are encountered. 
These are customarily set to zero, which is not the case here. 
In \cite{Shelton:2005cf} they were set to zero, in order for $f$ and $Q$ to be individually T-dual to 
$H$-flux from a 4-dimensional viewpoint.
Here we are in a totally different context, where we ask for coexistence of fluxes without a geometric 
dual. From a higher-dimensional viewpoint, a vanishing of these quantities is related to the compactness of 
the internal manifold \cite{SS}. However, in the present case all of them are derived quantities, 
decomposable into $f_{12}^3, B_{23}$ and $\beta^{12}$, and they appear because of the rotation of the bases 
that we used to construct the Courant algebroid. The underlying nilmanifold is still the one 
with structure constant $f_{12}^3$.

We conclude that the geometric deformation considered 
dresses the 3-dimensional nilmanifold with a large set of coexistent fluxes of all types.

\section{Discussion on flux coexistence and related issues}

The O(d,d) global symmetry of $(10{-}\text d)$-dimensional gauged supergravities inherited by T-duality led to
the introduction of general gaugings corresponding to so-called ``non-geometric'' fluxes \cite{Shelton:2005cf}. 
However, it still remains a puzzle whether these lower-dimensional supergravities have a 10-dimensional origin or, in 
other words, how they are obtained by dimensional reduction. Perhaps the most critical question in this discussion is whether the multiflux situations in four dimensions, expressed for example in terms of 
gauge algebras with general structure constants, are vacuous in ten. This may be reformulated as a 
question on the amount of fluxes that an internal space can admit or, posed differently, on whether all types 
of general fluxes can coexist in ten dimensions in a mathematically meaningful manner.

There are two possibilities related to flux coexistence. We begin with the less remarkable one. This amounts to 
starting with an appropriate nilmanifold, rich enough in geometric flux, and applying consecutive T-dualities to 
reach a situation with all types of fluxes. This is close in spirit to the conventional point of view on the 
subject that goes through duality transformations, and it is usually expressed in terms of the flux chain 
\be 
H_{ijk} \overset{T_i}\longleftrightarrow f_{jk}^{i} \overset{T_j}\longleftrightarrow Q_k^{ij} 
\overset{T_k}\longleftrightarrow R^{ijk}.
\label{chain}\ee
Clearly, this chain is the simplest possible one but very far from being the most general. Indeed, an 
arbitrary nilmanifold has multiple structure constants, and one may dualize, in some cases 
formally, along directions that have 
a different effect on each one. In order to illustrate this situation with an example, let us consider the 
6-dimensional nilmanifold with structure constants $f_{13}^{4},f_{14}^{6},f_{23}^{5},f_{25}^{6}$. 
This is a step-3 nilmanifold which was studied in the third ref.\ of \cite{nil}. The simple flux 
chain (\ref{chain}) is then enhanced to
\bea
  \left\{\begin{array}{c}  f_{13}^{4}\vspace{2pt}\\ f_{14}^{6}\vspace{2pt} \\ f_{23}^{5}\vspace{2pt} \\ f_{25}^{6} \end{array}\right\}
\quad \overset{T_6}\longleftrightarrow
\quad  \left\{\begin{array}{c}  f_{13}^{4}\vspace{2pt}\\ H_{146}\vspace{2pt} \\ f_{23}^{5}\vspace{2pt}\\ H_{256} \end{array}\right\}
\quad \overset{T_3}\longleftrightarrow
\quad  \left\{\begin{array}{c}  Q_{1}^{34}\vspace{2pt}\\ H_{146}\vspace{2pt} \\ Q_{2}^{35} \vspace{2pt}\\ H_{256} \end{array}\right\}
\quad \overset{T_1}\longleftrightarrow
\quad  \left\{\begin{array}{c}  R^{134}\vspace{2pt}\\ f_{46}^{1}\vspace{2pt} \\ Q_{2}^{35} \vspace{2pt}\\ H_{256} \end{array}\right\}.
\label{fHQR}
\eea
The rightmost entry indeed contains all types of fluxes, but it is obvious that it has a geometric dual at the leftmost entry. 
This happens because different fluxes penetrate different cycles of the manifold.
Therefore we will not delve into more details about this case.
More constructions along these lines appeared in \cite{Hassler:2013wsa}. Furthermore, it should be mentioned that both (\ref{chain}) 
and (\ref{fHQR}), although they can be understood from a four-dimensional viewpoint, they are less clear in ten dimensions. 
Indeed, the last dualities cannot be performed with the standard procedure of the Buscher rules because the corresponding isometries are 
broken in these cases. This problem may be overcome in the context of generalized T-duality, as in ref. \cite{Dall'Agata:2007sr}. 
However, as already mentioned, we are not working on such cases in the present paper and therefore this is not further discussed.

A more interesting construction was presented in the present paper. In particular, we chose to work with Dirac structures, which 
are subbundles of the Courant algebroid where the skew-symmetric Courant bracket is also associative. Therefore, Dirac structures
are the physically sensible subbundles of a general Courant algebroid. After classifying these structures and discussing how fluxes are introduced, we showed that it is possible to turn on multiple coexistent fluxes along 
the same cycles of a nilmanifold. Even in the simplest 3-dimensional case, the corresponding Courant algebroid 
contains the fluxes
\be
  \left\{\begin{array}{c} H_{123} \vspace{2pt} \\ f_{12}^{3} \vspace{2pt}\\ R^{123} \end{array}\right\} \quad+\quad
   \left\{\begin{array}{c} F_{13}^{1} \vspace{2pt} \\ \tilde F_{12}^{1}\vspace{2pt} \\\check F_{23}^{3}\vspace{2pt} \\ \tilde Q^{12}_{3} \vspace{2pt}\\
   Q^{13}_{1}\vspace{2pt}\\ Q^{23}_{2}\end{array}\right\}.
  \ee
The difference with the previous case is evident. In particular there is no T-duality, formal or not, that could 
geometrize this situation in the standard, non-generalized, sense of geometry. This provides a clear proof of principle 
for the existence of genuine multiflux generalized geometries. Although the question whether these multiflux geometries correspond to true string backgrounds is not addressed in this paper, our results clearly motivate further investigation. A closely related issue is 
that there is no a priori reason that the mathematical quantities we introduced in this paper are in one to one correspondence to 
the fluxes of ref. \cite{Shelton:2005cf}. Although such a relation can be expected due to previous work on the 
subject, such as refs. \cite{Dall'Agata:2007sr,Blumenhagen:2012pc,Andriot:2013xca} for example, a more precise treatment that would show such a 
correspondence is due.
  
Since Dirac structures are often associated to D-branes \cite{Grange:2006es,Satoshi,Jurco}, we expect that, 
building on relations described in \cite{Chatzistavrakidis:2013jqa}, our result may be translated into brane language and that coexistence of fluxes corresponds to bound states 
of non-standard or exotic branes \cite{Bergshoeff:2011se,deBoer:2012ma} which source these fluxes. 

The generalized geometric approach to unconventional cases 
of manifolds with fluxes is also appropriate to address quantization. 
This was emphasized in \cite{Szabo}, where an elegant phase-space point of view was suggested. From another standpoint 
the phase-space interpretation was also advocated in \cite{Chatzistavrakidis:2012yp} using matrix theory. 
More recently, the phase-space structure of $R$ flux backgrounds was also broadly discussed in \cite{Bakas:2013jwa}. 

Finally, cases with multiple coexistence of fluxes were recently studied in the context of 
asymmetric orbifolds, where the connection to 4-dimensional gauge supergravities was 
discussed \cite{Condeescu:2012sp}. It would be interesting to relate this approach to the one adopted 
in the present paper.

\paragraph{Acknowledgements.} 
We would like to thank F.F.~Gautason, D.~Mylonas, J.~Vysok\'y, S.~Watamura and especially P.~Schupp for discussions. This work was partially supported by the Deutsche Forschungsgemeinschaft grant LE 838/13.


\begin{thebibliography}{99}
\addtolength{\itemsep}{-3pt}
  
\bibitem{FC}
  M.~Gra\~na,
  Phys.\ Rept.\  {\bf 423} (2006) 91
  [hep-th/0509003];\\
  M.R.~Douglas and S.~Kachru,
  Rev.\ Mod.\ Phys.\  {\bf 79} (2007) 733
  [hep-th/0610102];\\
  R.~Blumenhagen, B.~K\"ors, D.~L\"ust and S.~Stieberger,\\
  Phys.\ Rept.\  {\bf 445} (2007) 1 
  [hep-th/0610327].
  
\bibitem{SS}
  J.~Scherk and J.H.~Schwarz,
  Nucl.\ Phys.\ B {\bf 153} (1979) 61;\\
  N.~Kaloper and R.C.~Myers,
  JHEP {\bf 9905} (1999) 010
  [hep-th/9901045];\\
  K.~Dasgupta, G.~Rajesh and S.~Sethi,
  JHEP {\bf 9908} (1999) 023
  [hep-th/9908088].
  
\bibitem{kstt}
  S.~Kachru, M.B.~Schulz, P.K.~Tripathy and S.P.~Trivedi,\\
  JHEP {\bf 0303} (2003) 061
  [hep-th/0211182];\\
  S.~Hellerman, J.~McGreevy and B.~Williams,
  JHEP {\bf 0401} (2004) 024
  [hep-th/0208174];\\
  A.~Dabholkar and C.~Hull,
  JHEP {\bf 0309} (2003) 054
  [hep-th/0210209];\\
  C.M.~Hull,
  JHEP {\bf 0510} (2005) 065
  [hep-th/0406102].

\bibitem{Shelton:2005cf}
  J.~Shelton, W.~Taylor and B.~Wecht,
  JHEP {\bf 0510} (2005) 085
  [hep-th/0508133].

\bibitem{Dabholkar:2005ve}
  A.~Dabholkar and C.~Hull,
  JHEP {\bf 0605} (2006) 009
  [hep-th/0512005].

  \bibitem{gcg1}
  N.~Hitchin,
  Quart.\ J.\ Math.\ Oxford Ser.\  {\bf 54} (2003) 281
  [math/0209099 [math-dg]].
  
\bibitem{gcg2}
  M.~Gualtieri,
  ``Generalized complex geometry,'' \\
  DPhil thesis,
  arXiv:math/0401221 [math.DG].
  
\bibitem{gcgsugra}
  M.~Gra\~na, R.~Minasian, M.~Petrini and A.~Tomasiello,\\
  JHEP {\bf 0511} (2005) 020
  [hep-th/0505212];\\
  M.~Gra\~na, R.~Minasian, M.~Petrini and D.~Waldram,\\
  JHEP {\bf 0904} (2009) 075
  [arXiv:0807.4527 [hep-th]];\\
  A.~Coimbra, C.~Strickland-Constable and D.~Waldram,\\
  JHEP {\bf 1111} (2011) 091
  [arXiv:1107.1733 [hep-th]].

\bibitem{dft}
G. Aldazabal, D. Marqu\'es and C. N\'u\~nez, \\
  Class.\ Quant.\ Grav.\  {\bf 30} (2013) 163001
  [arXiv:1305.1907 [hep-th]]; \\
D.S.~Berman and D.C.~Thompson,
  ``Duality Symmetric String and M-Theory,''
  arXiv:1306.2643 [hep-th]; \\
O. Hohm, D. L\"ust and B. Zwiebach,\\
 Fortsch. Phys. \textbf{61} (2013) 926 [arXiv:1309.2977 [hep-th]].


\bibitem{sixmanifolds}
G.R.~Cavalcanti and M.~Gualtieri,\\
J. Symplectic Geom. \textbf{2} (2004)  393
[arXiv:math/0404451 [math.DG]].
  
\bibitem{nil}
 C.M.~Hull and R.A.~Reid-Edwards,
  Fortsch.\ Phys.\  {\bf 57} (2009) 862
  [hep-th/0503114];\\
\dittostraight,~
  JHEP {\bf 0610} (2006) 086
  [hep-th/0603094];\\
  G.~Dall'Agata and N.~Prezas,
  JHEP {\bf 0510} (2005) 103
  [hep-th/0509052];\\
  M.~Gra\~na, R.~Minasian, M.~Petrini and A.~Tomasiello,\\
  JHEP {\bf 0705} (2007) 031
  [hep-th/0609124].

\bibitem{Dall'Agata:2007sr}
  G.~Dall'Agata, N.~Prezas, H.~Samtleben and M.~Trigiante,\\
  Nucl.\ Phys.\ B {\bf 799} (2008) 80
  [arXiv:0712.1026 [hep-th]];\\
  C.M.~Hull and R.A.~Reid-Edwards,\\
  JHEP {\bf 0909} (2009) 014
  [arXiv:0902.4032 [hep-th]];\\
  R.A.~Reid-Edwards,
  JHEP {\bf 0906} (2009) 085
  [arXiv:0904.0380 [hep-th]].

\bibitem{dirac}
 T.~Courant,
 Trans. Amer. Math. Soc., {\bf 319} (1990) 631-661.
  
\bibitem{Halmagyi:2008dr}
  N.~Halmagyi,
  JHEP {\bf 0807} (2008) 137
  [arXiv:0805.4571 [hep-th]];\\
 \dittostraight,~
  ``Non-geometric Backgrounds and the First Order String Sigma Model,''
  arXiv:0906.2891 [hep-th].

\bibitem{Blumenhagen:2012pc}
  R.~Blumenhagen, A.~Deser, E.~Plauschinn and F.~Rennecke,\\
Class.\ Quant.\ Grav.\  {\bf 29} (2012) 135004
  [arXiv:1202.4934 [hep-th]];\\
 \dittostraight,~ Fortsch.\ Phys.\  {\bf 60} (2012) 1217
  [arXiv:1205.1522 [hep-th]];\\
\dittostraight,~
  JHEP {\bf 1302} (2013) 122
  [arXiv:1211.0030 [hep-th]];\\
  R.~Blumenhagen, A.~Deser, E.~Plauschinn, F.~Rennecke and C.~Schmid,\\
 Fortsch. Phys. \textbf{61} (2013) 893 [arXiv:1304.2784 [hep-th]].

  \bibitem{Szabo}
 D.~Mylonas, P.~Schupp and R.J.~Szabo,\\
  JHEP {\bf 1209} (2012) 012 
  [arXiv:1207.0926 [hep-th]].
  
\bibitem{Lust:2012fp}
  D.~L\"ust,
  PoS CORFU {\bf 2011} (2011) 086
  [arXiv:1205.0100 [hep-th]].
  
  \bibitem{Roy}
D.~Roytenberg, ``Courant algebroids, derived brackets and even symplectic supermanifolds,'' Ph.D. thesis, arXiv:math/9910078 [math.DG];\\
\dittostraight,~ 	Lett. Math. Phys. {\bf 61} (2002) 123 [arXiv:math/0112152 [math.QA]]
  
\bibitem{wein}
Z.-J.~Liu, A.~Weinstein, and P.~Xu, 
J. Diff. Geom. {\bf 45} (1997) 547 [arXiv:dg-ga/9508013];\\
\dittostraight,~
Commun. Math. Phys. {\bf 192} (1998) 121 [arXiv:dg-ga/9611001].
  
\bibitem{severa}
H.~Bursztyn, M.~Crainic, and P.~\v{S}evera, 
Travaux math\'ematiques {\bf 16} (2005) 41.

\bibitem{DA09}
  D.~Andriot, R.~Minasian and M.~Petrini, \\
  JHEP {\bf 0912} (2009) 028
  [arXiv:0903.0633 [hep-th]].

\bibitem{ks}
 C.~Klim\v{c}\'ik and T.~Strobl,\\
 J. Geom. Phys. \textbf{43} (2002) 341 [arXiv:math/0104189 [math.SG]].

\bibitem{sw}
 P.~\v{S}evera and A.~Weinstein,\\
 Prog. Theor. Phys. Suppl. {\bf 144} (2001) 145 [arXiv:math/0107133 [math.SG]].

\bibitem{Aldazabal:2011nj}
  G.~Aldazabal, W.~Baron, D.~Marqu\'es and C.~N\'u\~nez,
  JHEP {\bf 1111} (2011) 052\\
  {}[Erratum-ibid.\  {\bf 1111} (2011) 109]
  [arXiv:1109.0290 [hep-th]].

  
  
\bibitem{Hassler:2013wsa}
  F.~Ha\ss{}ler and D.~L\"ust,
  JHEP {\bf 1307} (2013) 048
  [arXiv:1303.1413 [hep-th]].

\bibitem{Andriot:2013xca}
  D.~Andriot and A.~Betz,
  JHEP {\bf 1312} (2013) 083
  [arXiv:1306.4381 [hep-th]].

\bibitem{Grange:2006es}
  P.~Grange and S.~Sch\"afer-Nameki,
  Nucl.\ Phys.\ B {\bf 770} (2007) 123
  [hep-th/0609084].
  
  \bibitem{Satoshi}
  T.~Asakawa, S.~Sasa and S.~Watamura,\\
  JHEP {\bf 1210} (2012) 064
  [arXiv:1206.6964 [hep-th]].

\bibitem{Jurco}
  B.~Jur\v{c}o, P.~Schupp and J.~Vysok\'y,
  JHEP {\bf 1307} (2013) 126 
  [arXiv:1303.6096 [hep-th]].
  
\bibitem{Chatzistavrakidis:2013jqa}
  A.~Chatzistavrakidis, F.F.~Gautason, G.~Moutsopoulos and M.~Zagermann,\\
  Phys. Rev. D \textbf{89} (2014) 066004 [arXiv:1309.2653 [hep-th]].

\bibitem{Bergshoeff:2011se}
  E.A.~Bergshoeff, T.~Ort\'in and F.~Riccioni,\\
  Nucl.\ Phys.\ B {\bf 856} (2012) 210
  [arXiv:1109.4484 [hep-th]].
  
\bibitem{deBoer:2012ma}
  J.~de Boer and M.~Shigemori,\\
Phys.\ Rev.\ Lett.\  {\bf 104} (2010) 251603
  [arXiv:1004.2521 [hep-th]]; \\
\dittostraight,~  Phys. Rept. {\bf 532} (2013) 65
  [arXiv:1209.6056 [hep-th]].
  
\bibitem{Chatzistavrakidis:2012yp}
  A.~Chatzistavrakidis and L.~Jonke,\\
  Phys.\ Rev.\ D {\bf 85} (2012) 106013
  [arXiv:1202.4310 [hep-th]];\\
\dittostraight,~
  JHEP {\bf 1302} (2013) 040
  [arXiv:1207.6412 [hep-th]];\\
\dittostraight,~
  PoS Corfu {\bf 2012} (2013) 095
  [arXiv:1305.1864 [hep-th]].
  
\bibitem{Bakas:2013jwa}
  I.~Bakas and D.~L\"ust,
JHEP {\bf 1401} (2014) 17 [arXiv:1309.3172 [hep-th]].

\bibitem{Condeescu:2012sp}
  C.~Condeescu, I.~Florakis, C.~Kounnas and D.~L\"ust,\\
  JHEP {\bf 1310} (2013) 057
  [arXiv:1307.0999 [hep-th]].
  
\end{thebibliography}
\end{document}